\documentclass[conference]{IEEEtran}
\IEEEoverridecommandlockouts

\usepackage{amsmath,amssymb}
\usepackage{graphicx}
\usepackage{booktabs,tabularx,array,makecell}
\usepackage{cite}
\usepackage[final]{microtype}
\usepackage{xcolor}
\usepackage{tikz}
\usetikzlibrary{arrows.meta,calc,fit,positioning}
\usepackage{capt-of}
\usepackage{balance}

\PassOptionsToPackage{hyphens}{url}
\usepackage{xurl}
\PassOptionsToPackage{hyphens}{url}
\usepackage{xurl}
\renewcommand{\arraystretch}{1.08}
\newcolumntype{P}[1]{>{\raggedright\arraybackslash}p{#1}}
\def\BibTeX{{\rm B\kern-.05em{\sc i\kern-.025em b}\kern-.08em
    T\kern-.1667em\lower.7ex\hbox{E}\kern-.125emX}}

\begin{document}

\makeatletter
\def\bstctlcite#1{\@bsphack
  \@for\@citeb:=#1\do{\edef\@citeb{\expandafter\@firstofone\@citeb}%
    \if@filesw\immediate\write\@auxout{\string\citation{\@citeb}}\fi}%
  \@esphack}
\makeatother
\bstctlcite{IEEEexample:BSTcontrol}

\setlength{\columnsep}{0.30in}

\newcommand{\blue}[1]{\textcolor{blue}{#1}}
\newcommand{\red}[1]{{\textcolor{red}{#1}}}
\newcommand{\fa}[1]{\textcolor{magenta}{Fatemeh: #1}}
\newcommand{\ts}[1]{\textcolor{cyan}{Tolunay: #1}}
\newcommand{\rb}[1]{\textcolor{orange}{Ryan: #1}}
\newcommand{\ba}[1]{\textcolor{blue}{Basir:#1}}
\newcommand{\jb}[1]{\textcolor{purple}{Julia: #1}}
\renewcommand{\arraystretch}{1.1}

\title{
AtlasRAN: Timing-Aware Evaluation of Open-source 5G Platforms for Integrated Wireless Testbeds
\thanks{This work was supported, in part, at Clemson University by the State of South Carolina through funding for the Battelle Savannah River Alliance Workforce Development Program. It is endorsed by the NVIDIA Academic Grant Program. It is also funded by the National Science Foundation under Grant Numbers CNS-2202972, CNS-2318726, and CNS-2232048.}}

\author{
\IEEEauthorblockN{Ryan Barker}
\IEEEauthorblockA{\textit{Holcombe Department of Electrical}\\
\textit{and Computer Engineering}\\
\textit{Clemson University}\\
Clemson, SC, USA\\
rcbarke@clemson.edu}
\and
\IEEEauthorblockN{Tolunay Seyfi}
\IEEEauthorblockA{\textit{Holcombe Department of Electrical}\\
\textit{and Computer Engineering}\\
\textit{Clemson University}\\
Clemson, SC, USA\\
tseyfi@clemson.edu}
\and
\IEEEauthorblockN{Alireza Ebrahimi Dorcheh}
\IEEEauthorblockA{\textit{Holcombe Department of Electrical}\\
\textit{and Computer Engineering}\\
\textit{Clemson University}\\
Clemson, SC, USA\\
alireze@clemson.edu}
\and
\IEEEauthorblockN{Julia Boone}
\IEEEauthorblockA{\textit{Holcombe Department of Electrical}\\
\textit{and Computer Engineering}\\
\textit{Clemson University}\\
Clemson, SC, USA\\
jcboone@clemson.edu}
\and
\IEEEauthorblockN{Fatemeh Afghah}
\IEEEauthorblockA{\textit{Holcombe Department of Electrical}\\
\textit{and Computer Engineering}\\
\textit{Clemson University}\\
Clemson, SC, USA\\
fafghah@clemson.edu}
\and
\IEEEauthorblockN{Joseph Boccuzzi}
\IEEEauthorblockA{\textit{NVIDIA Corporation}\\
Santa Clara, CA, USA\\
jboccuzzi@nvidia.com}
}

\maketitle

\begin{abstract}
Open-source fifth-generation (5G) and Open Radio Access Network (O-RAN) experiments span simulation, host operating system (host-OS) emulation, software-defined radio hardware-in-the-loop testing, Open Radio Unit fronthaul deployments, wireless digital twins, and accelerator-backed radio access network (RAN) runtimes. These environments may expose similar interfaces while preserving different timing, input/output, synchronization, buffering, transport, and observability; \emph{functional compatibility is therefore not timing fidelity}.

This paper presents AtlasRAN, a claim-to-capability framework for deciding what an open-source 5G platform can credibly measure. It combines two reference paths, an execution-regime matrix, and a measurement-backed case study comparing OpenAirInterface (OAI) radio-frequency simulator (RFSim) with the Sionna Research Kit (Sionna-RK), which offloads low-density parity-check (LDPC) decoding to CUDA while retaining the surrounding OAI host-OS path. From one to six users, aggregate uplink goodput falls from 114.59 to 35.09~Mb/s for OAI and from 103.34 to 35.01~Mb/s for Sionna-RK. At six users, both RFSim real-time factors are approximately 0.55, despite estimated cumulative LDPC work of 1.09~ms per transport block for OAI and 0.37~ms for Sionna-RK. Timing-normalized service remains 60.7--63.6~Mb per emulated second across complete runs, while falling host and accelerator utilization is consistent with an under-fed decoder. A twelve-user run is retained only as failure-region evidence because end-of-test traffic summaries are absent. The practical takeaway is that integrated wireless testbeds, edge platforms, artificial-intelligence-enabled RANs, and digital twins should report timing discipline, transport path, memory movement, and observability as first-class experimental variables.
\end{abstract}

\begin{IEEEkeywords}
5G/6G, Open RAN (O-RAN), AI-RAN, wireless digital twins, testbeds, host-OS emulation, performance evaluation, CUDA-accelerated PHY
\end{IEEEkeywords}

\section{Introduction}
\label{sec:intro}

Open-source fifth-generation (5G) platforms are increasingly used as the experimental substrate for cloud--edge radio access networks (RANs), wireless digital twins, and shared tactical/civilian testbeds. They provide programmable protocol stacks, open control interfaces, and repeatable environments that are difficult to obtain in closed deployments. They also change what an experiment must prove. A 5G testbed is a coupled compute, transport, control, and observability system, so claims about throughput, latency, scalability, or closed-loop control depend on whether the platform preserves the target system's timing and input/output (I/O) behavior~\cite{kaltenberger2020_oai_democratizing,Colosseum2023,villa2025x5g}. In a tactical edge node that shares resources among radio processing, telemetry, control, and mission applications, a harness bottleneck misidentified as RAN capacity can lead to incorrect system dimensioning.

Common evaluation environments are nevertheless treated as interchangeable. A discrete-event simulator, a host operating system (host-OS) emulator, a software radio testbed, an open-fronthaul deployment, and a wireless digital twin may execute related protocol logic while exposing different scheduling, buffering, synchronization, transport, and telemetry behavior. Two studies can therefore appear to measure the same ``network performance'' while measuring different execution harnesses. AtlasRAN captures the central distinction: \emph{functional compatibility is not timing fidelity}. A platform may run the correct interfaces without preserving the time behavior required by the claim~\cite{lacava2023_ns_oran,robinson2024twinet,maxenti2025autoran}.

AtlasRAN uses Open Radio Access Network (O-RAN) disaggregation as the coordinate system that makes these assumptions visible. Centralized Unit (CU) and Distributed Unit (DU) placement, Open Radio Unit (O-RU) transport over Open Fronthaul (OFH), RAN Intelligent Controller (RIC) exposure, 5G Core (5GC) anchoring, and accelerator placement are all points at which timing, I/O, and observability enter a result. Artificial intelligence applied to the RAN (AI-RAN) adds another progression: a model may be trained offline, validated in a code-realistic twin, and executed near or inside the signal path, but each stage has a different timing contract~\cite{boccuzzi2025_gpu_accel_high_capacity_ai_ready,cohen2025_nvidia_ai_aerial}.

This paper makes four contributions:
\begin{itemize}
  \setlength{\itemsep}{0pt}
  \setlength{\parskip}{0pt}
  \item \textbf{Timing-aware framework:} two reference paths organize host-processor, radio, twin, and accelerator execution by the properties they preserve.
  \item \textbf{Claim-to-capability matrix:} research claims are mapped to required protocol depth, fronthaul realism, timing/I/O fidelity, control exposure, accelerator boundary, and observability.
  \item \textbf{Measurement-backed diagnosis:} a controlled ablation comparing the OpenAirInterface (OAI) radio-frequency simulator (RFSim) with the Sionna Research Kit (Sionna-RK) shows that a roughly threefold decoder-work reduction does not restore multi-user goodput when RFSim timing dilates and the pipeline is under-fed.
  \item \textbf{Open artifacts:} code, scripts, raw measurements, and figure inputs are released for reproduction and extension.\footnote{\url{https://github.com/rcbarke/atlasran}}
\end{itemize}

\section{Problem Statement \& Evaluation Contract}
\label{sec:problem_requirements}

Open-source 5G experimentation has a recurring validity problem: one execution regime is used to support claims about another. \emph{Functional compatibility} means that a platform runs the interfaces, procedures, or code paths required by a study. \emph{Timing fidelity} means that it also preserves the wall-clock timing, buffering, transport, synchronization, and I/O behavior needed for the claim. A host-OS emulator may support CU--DU execution, user equipment (UE) attachment, 5GC connectivity, and RIC telemetry while remaining unsuitable for claims about multi-user capacity, fronthaul timing, or slot deadlines.

AtlasRAN uses six execution-regime semantics. A \emph{discrete-event simulator} advances modeled time through an event queue. A \emph{host-OS emulator} runs RAN, user, and core processes on a general-purpose operating system while replacing the radio or fronthaul with sockets, queues, or channel abstractions. A \emph{software-defined radio (SDR) hardware-in-the-loop (HIL) testbed} closes the loop through a real radio-frequency (RF) front end and timing-coupled in-phase/quadrature (I/Q) stream, commonly with the full physical layer (PHY) in the host. An \emph{O-RU/OFH deployment} closes the low-PHY and RF endpoint in an O-RU and exposes O-RAN Working Group~4 (WG4) Split~7.2x transport, enhanced Common Public Radio Interface (eCPRI), synchronization, and radio-unit constraints. A \emph{wireless digital twin} couples network code to explicit scenario, propagation, and time semantics; it is not automatically cyber-physical unless bidirectional timing coupling is measured. An \emph{accelerator-backed runtime} makes central processing unit (CPU), graphics processing unit (GPU), data processing unit (DPU), network interface card (NIC), memory movement, and accelerator scheduling part of the result.

Three requirements follow. \textbf{R1: match platform class to claim.} Protocol attachment, near-real-time control, RF-facing performance, fronthaul timing, and in-path AI require different minimum realism. \textbf{R2: expose timing and transport.} Report synchronization, buffering, memory movement, transport, and any real-time-factor dilation. \textbf{R3: instrument the bottleneck.} Measurements must distinguish protocol, compute, inter-process communication (IPC), transport, and instrumentation limits.

Prior work provides open stacks, simulation, shared testbeds, twins, and accelerator runtimes~\cite{kaltenberger2020_oai_democratizing,srsran_project_overview_2026,lacava2023_ns_oran,Colosseum2023,villa2025x5g,robinson2024twinet,maxenti2025autoran,cohen2025_nvidia_ai_aerial}. AtlasRAN adds the missing claim-to-capability layer. It does not numerically ``correct'' a non-real-time trace into an over-the-air or deployment result. When timing fidelity fails, the defensible choices are to report performance jointly with the timing dilation, restrict the claim to functional behavior or a controlled ablation, or repeat the experiment in a regime that preserves the missing dependency.

\section{AtlasRAN Framework}
\label{sec:atlasran_framework}

AtlasRAN starts from the claim rather than the platform name. Protocol attachment may require only a functional stack and core path. Multi-user capacity, slot deadlines, fronthaul behavior, or online AI additionally require the relevant clock, transport, memory, and observability contract. The same software stack can therefore support one claim while being inappropriate for another.

\begin{figure*}[t]
\centering
  \includegraphics[width=0.9\textwidth]{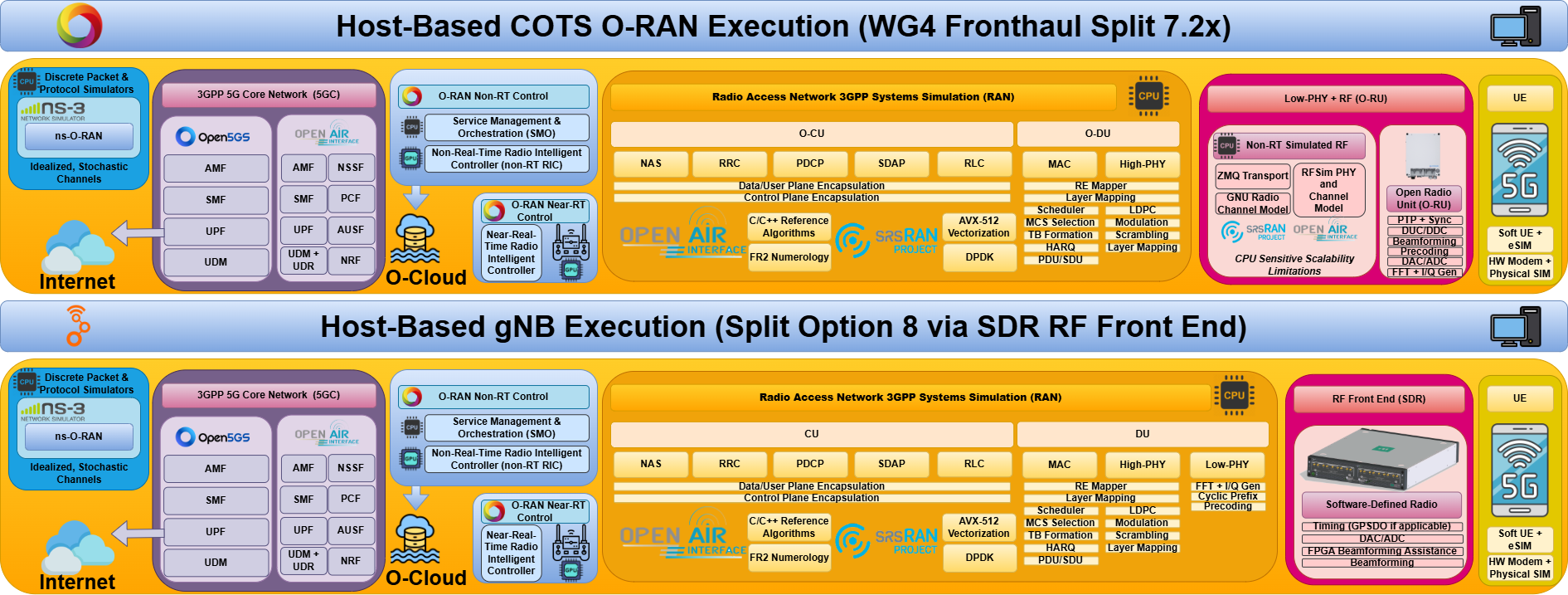}
  \caption{AtlasRAN CPU-centric path. Software radio, ordinary Split~8 SDR/HIL, and native Split~7.2x O-RU/OFH execution close the radio loop at different endpoints; that endpoint determines which timing, transport, synchronization, and interoperability constraints are measured rather than abstracted.}
  \label{fig:oran-host-os-emulation}
\end{figure*}

Figure~\ref{fig:oran-host-os-emulation} shows the CPU-centric path. Software radio emulation, Split~8 SDR/HIL, and native Split~7.2x O-RU/OFH execution share higher-layer components but close the radio loop at different endpoints. In software modes such as OAI RFSim, host scheduling, queues, sockets, and instrumentation determine wall-clock behavior~\cite{kaltenberger2020_oai_democratizing,oai_rfsimulator_readme}. Split~8 adds a real RF front end and time-domain I/Q transport while retaining the PHY in the host. Split~7.2x moves the low-PHY/RF endpoint into the O-RU, enabling frequency-domain I/Q transport with ethernet compression, and making fronthaul packet timing, clocking, radio capability matching, and O-DU/O-RU interoperability part of the experiment~\cite{srsran_cu_du_split_doc,srsran_oran_72_ru_guide,oai_oran_fhi72_tutorial_2024}.

\begin{figure}[htbp]
\centering
  \includegraphics[width=\columnwidth]{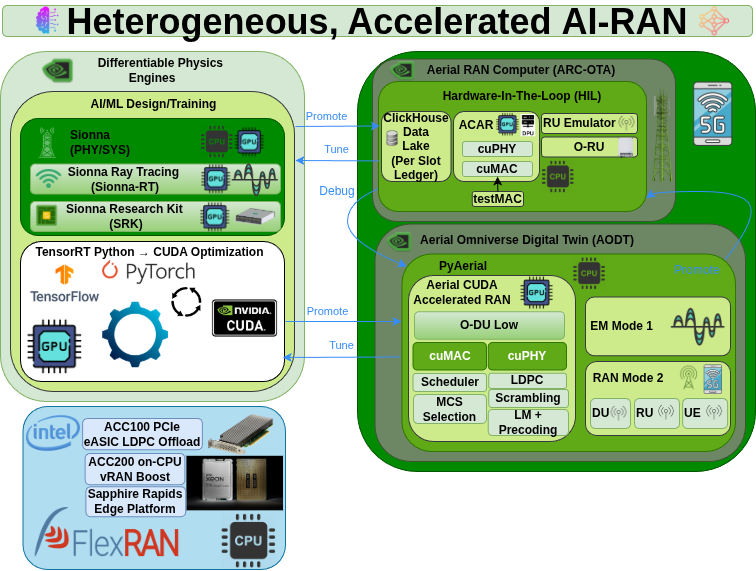}
  \caption{AtlasRAN accelerator/twin path for AI-RAN experimentation. Offline differentiable PHY/SYS tools, code-realistic twins, and deployment-grade accelerator runtimes form a promote--tune--debug loop. Moving along this path changes the experimental contract from differentiability and batch throughput to real-time memory movement, accelerator scheduling, fronthaul I/O, slot deadlines, and per-slot observability.}
  \label{fig:heterogeneous-ai-ran}
\end{figure}

Figure~\ref{fig:heterogeneous-ai-ran} shows the accelerator/twin path. Sionna and Sionna Ray Tracing (Sionna-RT) support offline differentiable PHY/system modeling and geometry-aware channel generation~\cite{hoydis2022_sionna,hoydis2023_sionna_rt}. Sionna-RK inserts selected graphics processing unit (GPU) functions, including low-density parity-check (LDPC) decoding, into an OAI path~\cite{sionna-rk}. NVIDIA Aerial Omniverse Digital Twin (AODT) and Aerial CUDA-Accelerated RAN (ACAR) connect code-realistic validation to accelerator-backed execution~\cite{nvidia2025_aerial_dt,boccuzzi2025_gpu_accel_high_capacity_ai_ready,cohen2025_nvidia_ai_aerial}. AtlasRAN distinguishes targeted kernel ablation, lookaside offload, and inline execution. In lookaside acceleration, the host invokes selected work and receives results across host--accelerator transactions; in inline acceleration, the full Layer~1 (L1) and fronthaul path remain on accelerator-side I/O~\cite{gadiyar2023_inline_vran,lo_schiavo_cloudric_mobicom24}. These are different experimental contracts, not interchangeable labels.

Table~\ref{tab:claim_capability_matrix} is the AtlasRAN platform ladder. It is organized by preserved contract rather than tool popularity. A reported user count must also state whether it represents modeled, registered, payload-active, telemetry-visible, RF-facing, or O-RU-backed users.

\begin{table*}[t]
\centering
\caption{AtlasRAN execution regimes, preserved contracts, defensible claims, and unsupported inferences.}
\label{tab:claim_capability_matrix}
\begingroup
\fontsize{7.10pt}{7.65pt}\selectfont
\setlength{\tabcolsep}{1.15pt}
\renewcommand{\arraystretch}{1.02}

\begin{tabularx}{0.965\textwidth}{
  @{}
  P{0.105\textwidth}
  P{0.150\textwidth}
  P{0.200\textwidth}
  @{\hspace{0.28in}}
  P{0.205\textwidth}
  >{\raggedright\arraybackslash}X
  @{}
}
\toprule
\textbf{Regime} &
\textbf{Representative platforms} &
\textbf{Preserved contract} &
\textbf{Defensible scaling claim} &
\textbf{What can go wrong / unsupported inference} \\
\midrule
Offline simulation / PHY-system modeling
&
ns-O-RAN; Sionna; Sionna-RT.
&
Event or batch time with explicit traffic, scheduler, channel, or ray-tracing models; no live RF or wall-clock claim.
&
Large-$N$ topology, channel, policy, or dataset scale. User count is model scale, not live-stack capacity.
&
Event-queue complexity, abstraction error, batch-memory limits, or modeled scheduler assumptions can dominate. Simulated user count should not be reported as full-stack RAN capacity. \\

\midrule

Protocol attachment / registration
&
OAI RFSim attach tests; srsRAN software-radio tests; emulated RIC control tests.
&
Control-plane procedures, 5GC path, and per-user logs; no concurrent payload-capacity assumption.
&
Attachment, registration, and control-plane scale. Registered and payload-active users must be reported separately.
&
Protocol timers, user orchestration, subscriber/core queues, and container scheduling can dominate. Registered users are not concurrent data-plane users. \\

\midrule

Host-OS payload emulation
&
OAI RFSim throughput; Sionna-RK over OAI RFSim; srsRAN software sample transport.
&
RAN--user--5GC payload path through software radio/sample transport; real-time factor, socket health, dropped flows, and log completeness measured.
&
Small-$N$ payload and controlled ablation claims. Here, $N\!\in\!\{1,3,6\}$ is the clean evidence region; $N\!=\!12$ is a failure marker.
&
Process-to-process backpressure, sample sockets, transport buffering, host scheduling, timing dilation, and telemetry starvation. Emulated goodput is not over-the-air, O-RU, or deployment capacity. \\

\midrule

Near-real-time control / telemetry
&
OAI or srsRAN with RIC integration; e2sim; ns-O-RAN for simulation-scale control.
&
Repeatable telemetry/control exposure, key-performance-indicator cadence, and controllable RAN parameters.
&
Control-application logic scale, telemetry cadence, policy stability, and action repeatability. User count measures control-loop load, not necessarily radio capacity.
&
Serialization, sampling interval, queues, transport, logging, and subscriber backpressure can dominate. Convergence under emulation is not deployed near-real-time latency. \\

\midrule

Software / wireless digital twin
&
AODT-style twin; Sionna-RT-assisted twin; software radio/user endpoints; Colosseum/OpenRAN Gym when used repeatably.
&
Explicit scenario state, propagation or software-radio model, slot/event progression, and repeatable state capture.
&
Software-twin scale, model promotion, debugging, and repeatable stress testing; not automatically fronthaul-faithful capacity.
&
Twin scheduling, radio/user-model fidelity, GPU batch scheduling, a missing NIC-to-GPU fronthaul path, or omitted clock/switch behavior can dominate. Twin goodput is not deployed O-RU throughput. \\

\midrule

SDR hardware-in-the-loop
&
OAI-SDR; srsRAN-SDR; Colosseum/OpenRAN Gym; X5G; shared integration testbeds.
&
Real RF front end with timing-coupled I/Q stream; the full PHY usually remains in the host in ordinary Split~8.
&
RF-facing throughput, attachment, mobility, and scheduling for the measured radio, channel, bandwidth, user mix, and clocking setup.
&
Calibration, front-end linearity, radio underruns/overruns, sample transport, host PHY load, clock drift, and commercial/software-user limits can dominate. Split~8 is not native OFH evidence. \\

\midrule

O-RU/OFH end-to-end deployment
&
OAI or srsRAN with a compatible O-RU; dedicated O-RU interoperability testbeds.
&
Split~7.2x OFH/eCPRI, radio capability matching, Precision Time Protocol (PTP) or Global Positioning System (GPS)-disciplined-oscillator synchronization, fronthaul switching, deadline counters, and user-plane traffic.
&
Fronthaul timing, O-RU interoperability, and end-to-end capacity for the tested radio, clock tree, switch, bandwidth, multiple-input multiple-output mode, and O-DU configuration.
&
Clock lock, packet jitter/pacing, fronthaul network configuration, NIC path, CPU isolation, low-PHY deadlines, radio limits, random access, and sounding can dominate. Software-radio results do not prove native OFH behavior. \\

\midrule

Accelerator-backed RAN runtime
&
Sionna-RK LDPC ablation; FlexRAN/CloudRIC lookaside paths; ACAR/Aerial inline GPU/DPU runtime.
&
Explicit acceleration boundary: targeted kernel, selected-function lookaside, or full-L1 inline; memory movement and fronthaul/NIC path reported.
&
Kernel-offload, CPU-headroom, AI-on-RAN, or inline L1/fronthaul claims. End-to-end capacity also requires live RF/fronthaul timing.
&
Host--accelerator transactions, batching, memory pressure, direct-memory-access/copy overheads, NIC--GPU configuration, GPU scheduling, and telemetry can dominate. LDPC-only speedup does not prove multi-user RAN capacity. \\

\bottomrule
\end{tabularx}
\endgroup
\end{table*}

The matrix provides direct use guidance. Use simulation for topology, channel, policy, or dataset scale without wall-clock claims; host-OS emulation for attachment, protocol regression, control-application iteration, logging, and controlled ablations; SDR/HIL for RF and clocking claims; O-RU/OFH for Split~7.2x transport and interoperability; and timing-faithful twins or inline runtimes when the claim depends on in-path AI, full-L1 acceleration, or NIC--GPU movement. The least complex valid platform is the one that preserves the strongest timing and I/O dependency in the claim. The case study below exercises the host-OS payload-emulation row, not a universal capacity benchmark.

\section{CU--DU Case Study Methodology}
\label{sec:methodology}

\begin{figure*}[t]
\centering
\includegraphics[width=0.90\textwidth]{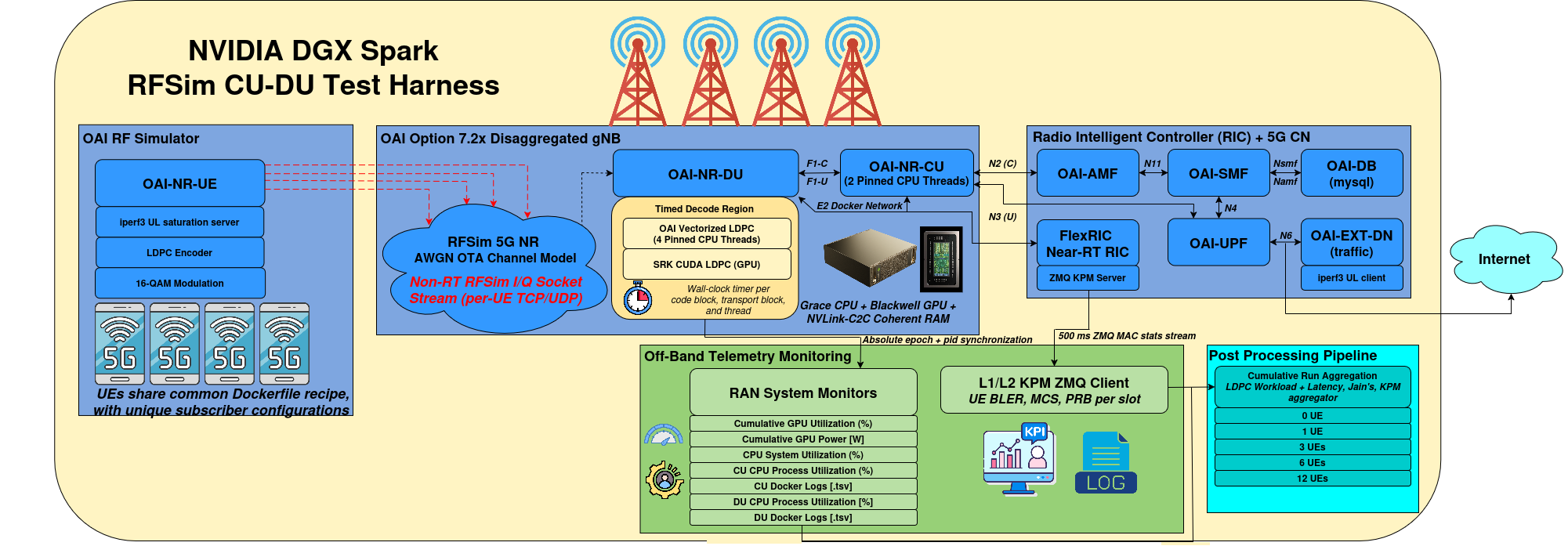}
\captionof{figure}{Uplink TCP saturation case-study abstraction. The two configurations share the CU, core, RFSim, traffic, and monitoring path; only the LDPC decode region changes. This is a host-OS ablation, not over-the-air, native open-fronthaul, or full-L1 inline evidence.}
\label{fig:cu_du_harness}
\end{figure*}

We compare baseline OAI RFSim with Sionna-RK on the common harness in Fig.~\ref{fig:cu_du_harness}. Both configurations run an OAI 5G base station (gNB) with the standardized F1 interface between the CU and DU on one NVIDIA DGX Spark. RFSim replaces the physical radio and open-fronthaul endpoint with a software channel and sample-socket path. The CU and DU are pinned to disjoint CPU sets; remaining cores support user containers, traffic generation, monitoring, and post-processing.

The compute change is deliberately narrow. In OAI, Physical Uplink Shared Channel (PUSCH) LDPC code blocks are dispatched across four CPU workers, each using single-instruction multiple-data check-node and variable-node kernels. Sionna-RK replaces that decode region with a CUDA min-sum decoder and packed hard decisions on one CUDA stream. Table~\ref{tab:cu-du-kpis} therefore compares an estimated four-worker CPU core-time with one CUDA decode call. It does not compare OAI against a full-L1 inline runtime; the surrounding RFSim, host, and F1 paths remain common~\cite{sionna-rk,gadiyar2023_inline_vran}.

For each $N\!\in\!\{1,3,6,12\}$, we run a 60~s saturated uplink Transmission Control Protocol workload using one \texttt{iperf3} flow per user and compute aggregate goodput at the data-network receiver. We sample per-user goodput, Jain's fairness index (where one denotes equal per-user goodput), CU/DU and host CPU use, Sionna-RK GPU use and power, LDPC timing, RFSim slot progression, and telemetry health. ZeroMQ (ZMQ) carries external key-performance-indicator data to a RIC-application-style subscriber; it is instrumentation, not srsRAN's virtual-radio mode. Full raw logs and 95th-percentile summaries are retained in the artifact.

A monitor prepends wall-clock epochs to DU slot logs. For numerology $\mu=1$, the slot duration is $\Delta_{\mathrm{slot}}=0.5$~ms. We define the real-time factor (RTF) and its dilation $D_{\mathrm{RTF}}$ as
\begin{equation}
\mathrm{RTF}=\frac{(s(t)-s_0)\Delta_{\mathrm{slot}}}{t-t_0}, \qquad D_{\mathrm{RTF}}=\frac{1}{\mathrm{RTF}},
\label{eq:rtf}
\end{equation}
where $s(t)$ is the DU slot index. RTF equals one for wall-clock progression, exceeds one when RFSim runs ahead, and falls below one when emulated RAN time lags. $D_{\mathrm{RTF}}$ is wall-clock seconds per emulated second. The experiment asks whether lowering decoder work restores multi-user scaling under the same host-OS harness; it makes no over-the-air, O-RU interoperability, or universal user-limit claim.

\begin{table*}[t]
\caption{Core KPIs for the CU--DU load study. CPU/LDPC/GPU cells report
mean/p95 from 0.5~s samples over 60~s; goodput, RTF, and Jain's $J$ are
run-level. CPU is core-equivalent (100\% is one logical core).
``LDPC/thread'' is decoder-call wall time; ``LDPC cum.'' is
per-transport-block decoder work (OAI: four-worker core-time; SRK:
one CUDA call).}
\label{tab:cu-du-kpis}
\centering
\fontsize{7.25pt}{7.9pt}\selectfont
\setlength{\tabcolsep}{1.2pt}
\renewcommand{\arraystretch}{1.08}
\resizebox{\textwidth}{!}{%
\begin{tabular}{c | rr | rr | rr | rr | rr | rr | rr | rr | rr | rr}
\toprule
& \multicolumn{2}{c|}{Goodput (Mb/s)}
& \multicolumn{2}{c|}{Per-UE (Mb/s)}
& \multicolumn{2}{c|}{RTF}
& \multicolumn{2}{c|}{Jain $J$}
& \multicolumn{2}{c|}{DU CPU (\%)}
& \multicolumn{2}{c|}{CU CPU (\%)}
& \multicolumn{2}{c|}{Host CPU (\%)}
& \multicolumn{2}{c|}{LDPC/thread ($\mu$s)}
& \multicolumn{2}{c|}{LDPC cum. ($\mu$s)}
& \multicolumn{2}{c}{SRK GPU (util\% / W)} \\
\cmidrule(lr){2-3}
\cmidrule(lr){4-5}
\cmidrule(lr){6-7}
\cmidrule(lr){8-9}
\cmidrule(lr){10-11}
\cmidrule(lr){12-13}
\cmidrule(lr){14-15}
\cmidrule(lr){16-17}
\cmidrule(lr){18-19}
\cmidrule(lr){20-21}
$N$
& OAI & SRK
& OAI & SRK
& OAI & SRK
& OAI & SRK
& OAI & SRK
& OAI & SRK
& OAI & SRK
& OAI & SRK
& OAI & SRK
& util & power \\
\midrule
0
& -- & --
& -- & --
& 3.741 & 3.693
& -- & --
& 210.16/213.27 & 212.87/216.09
& 1.71/2.29 & 1.87/2.84
& 305.6/336.6 & 301.6/340.0
& 79.87/91.10 & --
& 79.87/91.10 & --
& 1.29/3.70 & 11.48/12.11 \\
1
& 114.59 & 103.34
& 114.59 & 103.34
& 1.844 & 1.703
& 1.000000 & 1.000000
& 324.79/326.77 & 349.31/351.64
& 22.98/24.43 & 20.73/22.54
& 534.6/561.6 & 557.2/594.8
& 255.86/262.09 & 316.02/350.83
& 1016.74/1048.36 & 316.02/350.83
& 44.85/47.00 & 38.48/39.11 \\
3
& 65.21 & 66.44
& 21.74 & 22.15
& 1.046 & 1.061
& 0.999997 & 1.000000
& 289.93/291.73 & 329.73/339.57
& 16.80/17.68 & 17.21/18.11
& 502.2/527.4 & 585.0/636.4
& 262.60/268.36 & 316.90/347.89
& 1042.06/1073.45 & 316.90/347.89
& 29.28/34.00 & 29.73/30.54 \\
6
& 35.09 & 35.01
& 5.85 & 5.84
& 0.552 & 0.551
& 0.999484 & 0.999930
& 205.54/207.18 & 231.60/244.69
& 9.02/10.63 & 9.77/10.75
& 391.6/428.6 & 484.6/521.6
& 275.27/286.05 & 369.47/392.66
& 1087.42/1144.18 & 369.47/392.66
& 19.34/22.00 & 22.25/23.14 \\
12$^\dagger$
& 16.35 & 16.15
& 1.36 & 1.35
& 0.285 & 0.282
& 0.999997 & 0.999998
& 162.38/164.92 & 172.91/176.62
& 5.69/6.89 & 5.80/6.78
& 338.2/359.0 & 389.2/414.0
& 273.18/293.63 & 345.17/373.05
& 1063.82/1174.53 & 345.17/373.05
& 9.45/11.00 & 17.08/17.60 \\
\bottomrule
\end{tabular}}

\vspace{0.2em}
\begin{minipage}{0.96\textwidth}
\scriptsize
$^\dagger$At $N\!=\!12$, all traffic files parse but end-of-test
summaries are absent, and telemetry is degraded. This row is failure-region evidence; it is not a clean steady-state capacity point.
\end{minipage}
\end{table*}

\section{Results: Timing-Dilated Scaling}
\label{sec:results}

\begin{figure}[t]
\centering
  \includegraphics[width=\columnwidth]{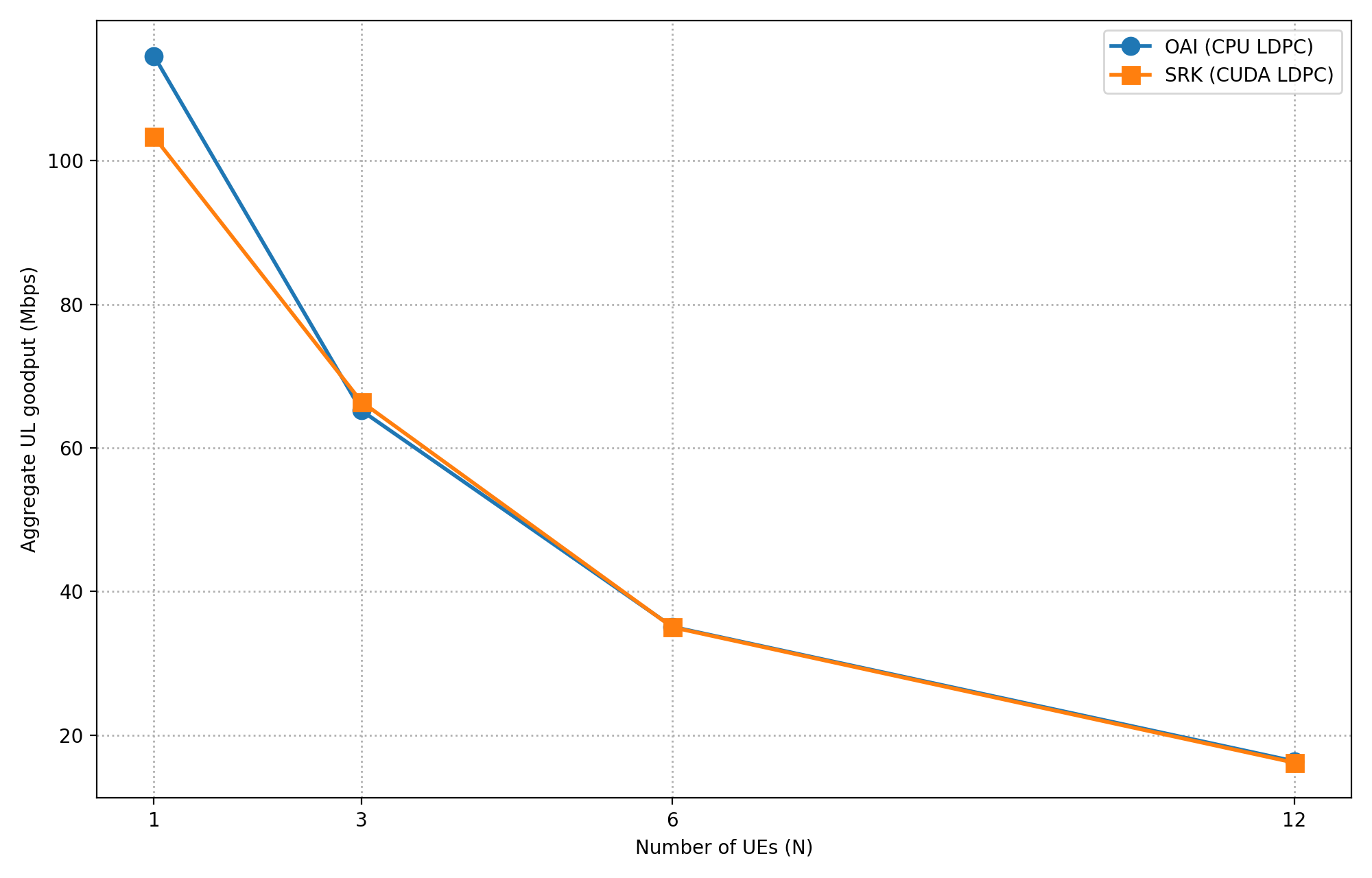}
  \caption{Aggregate wall-clock uplink goodput for OAI RFSim with CPU LDPC and Sionna-RK with CUDA LDPC. Solid segments contain complete runs; dashed segments and open markers lead to $N\!=\!12$, a degraded failure-region point rather than a capacity datum or fit anchor.}
  \label{fig:goodput_scaling}
\end{figure}

Figure~\ref{fig:goodput_scaling} reports wall-clock uplink goodput. The configured link has a nominal raw bound of $R_{\mathrm{raw}}\!\approx\!142.46$~Mb per emulated radio second (106 resource blocks, 30~kHz spacing, 14 symbols/slot, and 16-state quadrature amplitude modulation). At one user, OAI/Sionna-RK records 114.59/103.34~Mb/s; because RTF differs from one, those wall-clock rates are not directly comparable to the bound. The runs are complete through six users, where both rates are about 35~Mb/s; $N\!=\!12$ is only a failure marker.

The timing measurements expose the shared clamp. RTF falls from 3.74/3.69 at idle to 1.84/1.70 at one user and 0.552/0.551 at six users for OAI/Sionna-RK. From three to six users, OAI RTF falls 47.2\% while goodput falls 46.2\%; Sionna-RK RTF falls 48.1\% while goodput falls 47.3\%. These co-moving reductions occur in runs with complete traffic summaries. At six users, the two goodputs differ by only 0.22\%, and both RTFs are about 0.55, even though the LDPC work differs by 2.94$\times$. The shared host-OS timing path therefore best explains the measured clamp; this experiment does not isolate one socket or scheduler mechanism.

A timing-normalized diagnostic separates protocol-time service from wall-clock delivery. Let $T_{\mathrm{wall}}$ be receiver goodput per wall-clock second and $T_{\mathrm{emu}}$ payload per emulated radio second:
\begin{equation}
T_{\mathrm{emu}}=\frac{T_{\mathrm{wall}}}{\mathrm{RTF}}.
\label{eq:timing_normalized_goodput}
\end{equation}
For OAI, $T_{\mathrm{emu}}$ is 62.1, 62.4, and 63.6~Mb per emulated second at one, three, and six users; Sionna-RK yields 60.7, 62.6, and 63.5. Thus, service per emulated radio second remains nearly constant while service per wall-clock second falls. Equation~\eqref{eq:timing_normalized_goodput} is a diagnostic normalization, not a conversion to over-the-air or deployment capacity.

Fairness and utilization rule out two alternative explanations in the clean region. At six users, Jain's fairness index is 0.999484 for OAI and 0.999930 for Sionna-RK, so users slow together rather than one user capturing the scheduler. From one to six users, OAI DU CPU falls from 324.79\% to 205.54\%, Sionna-RK DU CPU from 349.31\% to 231.60\%, and Sionna-RK GPU utilization from 44.85\% to 19.34\%. Host-wide CPU also falls. A sustained decoder or host-capacity bottleneck would ordinarily keep the limiting resource near saturation; here, resource use falls as less useful work arrives per wall-clock second.

The LDPC columns isolate the compute ablation. OAI requires 1.02--1.09~ms of estimated cumulative CPU work per transport block, while Sionna-RK requires 0.32--0.37~ms. Let $U_{\mathrm{LDPC}}$ denote decoder utilization, $\lambda_{\mathrm{TB}}$ useful transport blocks delivered per wall-clock second, and $t^{\mathrm{cum}}_{\mathrm{LDPC}}$ cumulative decoder work per block. Then
\begin{equation}
U_{\mathrm{LDPC}}\propto \lambda_{\mathrm{TB}}t^{\mathrm{cum}}_{\mathrm{LDPC}}.
\label{eq:underfeeding}
\end{equation}
Sionna-RK reduces $t^{\mathrm{cum}}_{\mathrm{LDPC}}$, but socket backpressure and timing dilation reduce $\lambda_{\mathrm{TB}}$; the accelerator is consequently under-fed. This does not diminish the decoder speedup. It limits the defensible claim to a decoder-kernel ablation, not full-L1 or multi-user RAN capacity~\cite{gadiyar2023_inline_vran}.

The twelve-user run marks where the measurement surface fails. RTF is approximately 0.28, so one emulated second requires about 3.5 wall-clock seconds, and end-of-test traffic summaries are missing. Those artifacts reveal IPC collapse, timing dilation, and observability loss, but they must not anchor a scaling fit or platform-wide capacity claim. RFSim remains valuable for attachment, protocol integration, instrumentation, and controlled ablations; the result shows why its stressed goodput must not be relabeled as over-the-air, O-RU/OFH, or deployment-grade deadline behavior.

\section{Discussion}
\label{sec:discussion}

AtlasRAN converts the case-study diagnosis into a reporting discipline. Authors should state the execution regime; timing discipline, including synchronization, deadlines, jitter, and RTF; transport and memory path, including whether acceleration is targeted, lookaside, or inline; and observability, including cadence, dropped logs, and incomplete summaries. When RTF or instrumentation fails, reporting throughput alone is insufficient. The result should be paired with dilation, narrowed to the valid ablation, or promoted to a more faithful regime. There is no defensible scalar that turns a dilated host-OS trace into native fronthaul capacity.
For timing-dilated emulation, report both wall-clock goodput and timing-normalized service. The former captures what a co-located application or operator receives per real second; the latter shows how much payload the emulated protocol delivers per emulated radio second. They answer different questions and should never be substituted without the accompanying RTF.

For AI-RAN, the same rule applies to data and control. Offline Sionna/Sionna-RT studies are appropriate for differentiable models and dataset scale; a code-realistic twin is appropriate for repeatable promotion and debugging; and an inline runtime is required when the claim depends on fronthaul I/O, slot deadlines, or in-path inference~\cite{hoydis2022_sionna,hoydis2023_sionna_rt,nvidia2025_aerial_dt,boccuzzi2025_gpu_accel_high_capacity_ai_ready}. Otherwise, training and validation traces can encode host scheduling, socket backpressure, or telemetry stalls rather than the intended radio dynamics.

The case study is intentionally narrow: one CPU--GPU host, one OAI RFSim-derived family, one uplink TCP workload, one CU--DU split, and one accelerator variant. It establishes neither a universal user limit nor a global ranking of OAI, RFSim, Sionna-RK, or GPU acceleration. Its value is diagnostic: lowering decoder work by roughly threefold does not recover end-to-end scaling when the surrounding harness no longer delivers work at the required wall-clock rate. Future validation should repeat the same controlled ablation with SDR/HIL, native O-RU/OFH, and inline accelerator paths.

\section{Conclusion}
\label{sec:conclusion}

AtlasRAN treats platform choice as part of experimental method. Open-source 5G environments must be compared not only by interface support, but also by timing, transport, synchronization, memory movement, and observability. In the CU--DU study, CUDA LDPC offload reduces decoder work, yet multi-user goodput follows RFSim timing dilation while host and accelerator utilization fall. The result is a harness-limited ablation, not an OAI, Sionna-RK, or deployment-capacity limit. For tactical testbeds, wireless twins, and AI-RAN systems, the least complex valid platform is the one that preserves the timing and I/O dependency of the claim.

\balance
\bibliographystyle{IEEEtran}
\bibliography{References}

\end{document}